\documentclass[pra,superscriptaddress,twocolumn]{revtex4-1}
\usepackage{amsfonts}
\usepackage{amsmath}
\usepackage{amsthm}
\usepackage{mathrsfs}
\usepackage{amscd}
\usepackage{amssymb}
\usepackage{subfigure}
\usepackage{amsxtra}
\usepackage{bm}           
\usepackage{bbm}
\usepackage{graphicx}
\usepackage{epstopdf}

\usepackage[colorlinks]{hyperref}

\newcommand{\bra}[1]{\ensuremath{\langle#1|}}
\newcommand{\ket}[1]{\ensuremath{\left|#1\right\rangle}}

\newcommand{\mean}[1]{\ensuremath{\left\langle #1 \right\rangle}}

\newcommand{\dg}{\dagger}

\newcommand{\tr}{\ensuremath{{\rm tr}}}

\newcommand*\dd{\mathrm{d}}

\begin{document}
	
\author{Alexandre Roulet}
\affiliation{Centre for Quantum Technologies, National University of Singapore, 3 Science Drive 2, Singapore 117543, Singapore}

\author{Stefan Nimmrichter}
\affiliation{Centre for Quantum Technologies, National University of Singapore, 3 Science Drive 2, Singapore 117543, Singapore}

\author{Juan Miguel Arrazola}
\affiliation{Centre for Quantum Technologies, National University of Singapore, 3 Science Drive 2, Singapore 117543, Singapore}

\author{Stella Seah}
\affiliation{Department of Physics, National University of Singapore, 2 Science Drive 3, Singapore 117542, Singapore}

\author{Valerio Scarani}
\affiliation{Centre for Quantum Technologies, National University of Singapore, 3 Science Drive 2, Singapore 117543, Singapore}
\affiliation{Department of Physics, National University of Singapore, 2 Science Drive 3, Singapore 117542, Singapore}

\title{Autonomous Rotor Heat Engine}
\begin{abstract}
The triumph of heat engines is their ability to convert the disordered energy of thermal sources into useful mechanical motion. In recent years, much effort has been devoted to generalizing thermodynamic notions to the quantum regime, partly motivated by the promise of surpassing classical heat engines. Here, we instead adopt a bottom-up approach: we propose a realistic autonomous heat engine that can serve as a testbed for quantum effects in the context of thermodynamics. Our model draws inspiration from actual piston engines and is built from closed-system Hamiltonians and weak bath coupling terms. We analytically derive the performance of the engine in the classical regime via a set of nonlinear Langevin equations. In the quantum case, we perform numerical simulations of the master equation. Finally, we perform a dynamic and thermodynamic analysis of the engine's behaviour for several parameter regimes in both the classical and quantum case, and find that the latter exhibits a consistently lower efficiency due to additional noise.
\end{abstract}
\maketitle

\section{Introduction} 
Historically, the initial goal of thermodynamics was to understand how to convert heat into useful mechanical motion, and it was only once this goal was achieved with the rise of steam engines that there was an interest in exploring the theoretical limitations to their efficiency~\cite{carnot1824}. As our ability to control quantum systems progresses, it has now become interesting to study thermal machines where quantum effects are relevant. However, in this case, the historical order has been reversed, with earlier work focusing on the thermodynamic limitations of quantum machines~\cite{kieu2004,rezek2006,horodecki2013}. More recently, new proposals for thermal machines that do not require external sources of work have been made. 
These include absorption refrigerators \cite{linden2010small,kosloff2012,venturelli2013minimal,correa2014quantum,Maslennikov2017}, heat engines \cite{mari2015quantum,brunner2012virtual,teo2016converting,Bissbort2016minimalistic,serra2016mechanical,rossnagel2016single}, as well as thermoelectric devices converting between temperature bias and electrical current or voltage bias \cite{Sanchez2011,Thierschmann2015,Koski2015,Marchegiani2016,Alicki2016}. 
Additionally, there has been significant interest in determining the extent to which quantum effects such as squeezing or coherence may help surpass classical limits such as the Carnot efficiency~\cite{scully2003,rossnagel2014,zhang2014,jaramillo2016quantum,gardas2015thermodynamic,ng2016surpassing}, or complicate classical notions such as work~\cite{roncaglia2014,hanggi2016}.

In this work, we consider the problem of designing a quantum heat engine that achieves its goal of converting heat into the useful mechanical motion of a system. In particular, our goal is to devise a self-contained engine that autonomously converts heat from a thermal bath into motion in a single rotational degree of freedom. We choose to study a rotor to benefit from the useful features of rotational motion: it is inherently periodic, it can in principle be coupled through other systems such as gears and pistons into other types of motion, and it can be used to drive electric generators. Additionally, unlike the motion of oscillators, rotational motion has a meaningful sense of directionality~\cite{wang2013} through the distinction between clockwise and counter-clockwise rotation. The rotation frequency is not upper bounded, as would be the case in finite dimensional systems, and it unambiguously displays the amount of useful energy stored in the rotor degree of freedom.

\begin{figure}
\begin{center}
\includegraphics[width=0.85\columnwidth]{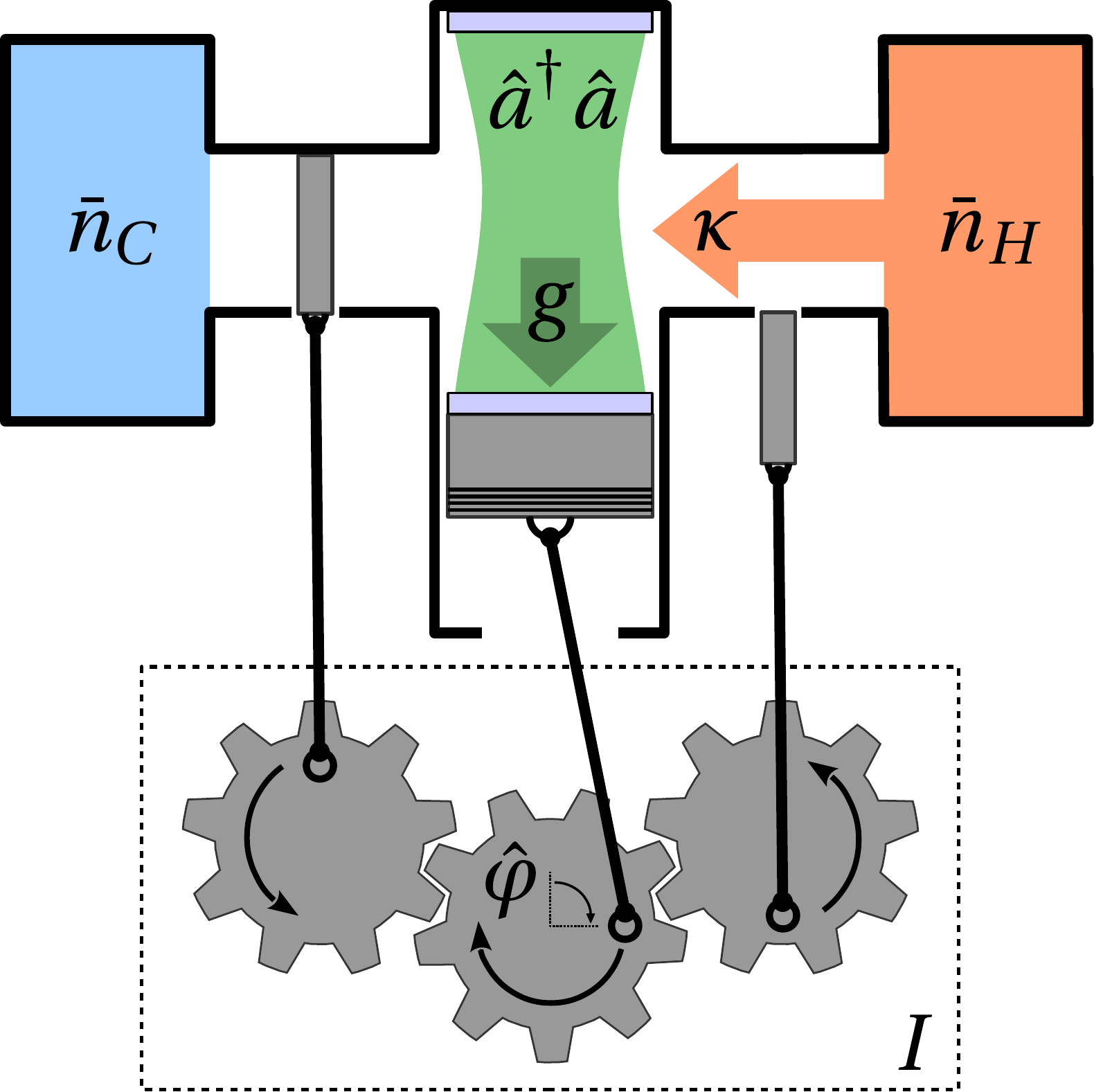}
\caption{Autonomous rotor heat engine. A harmonic mode is pushing down a piston attached to a rotor through radiation pressure. Concurrently, the angular position $\varphi$ of the rotor (defined relative to the upper turning point) modulates the coupling of the mode to baths at respective occupations $\bar{n}_H$ and $\bar{n}_C$. This leads to a preferred clockwise motion of the rotor. Note that the specific implementation of the valves depicted here realizes exactly the modulation functions \eqref{eq:modFunc}. The terms $\kappa$, $g$, and $I$ denote the bath thermalization rate, the torque per excitation in the mode, and the moment of inertia, respectively.}\label{fig:engine}
\end{center}
\end{figure}

We design a simple opto-mechanical engine where the rotor is coupled to a single harmonic mode through radiation pressure, which is in turn coupled linearly to two baths at different temperatures. In analogy to the hot gas pushing down the piston of an actual car engine, the mode serves as the working medium transferring heat from the hot to the cold bath in sync with the angular motion of the rotor. We characterize analytically the engine's operation in the classical regime and we find that it functions as desired. We then describe how these equations of motion can be solved, analytically in the classical case and numerically in the quantum regime, and study the dynamics for different values of the relevant parameters, focusing on the transient behaviour of the engine as it accelerates from rest. We then perform a thermodynamic analysis of the engine by computing the work output, heat input and efficiency of the engine in both the classical and quantum regimes. We conclude by discussing the significance of the various engine parameters and the role of quantum effects.

\section{Rotor heat engine}
\emph{Design.--} The guiding principle in our design of the rotor heat engine is that it should unambiguously achieve its intended goal of converting heat into useful mechanical motion of the rotor. To be precise, we demand that (i) the engine is autonomous, (ii) the rotor draws energy exclusively from a thermal source, and (iii) the rotor undergoes useful directional motion, i.e.~it has a well-defined angular momentum increasing with time. We allow for an initialization of the rotor at a well localized angle, but do not permit the use of external control fields and time-dependent Hamiltonians for the dynamics, contrary to previous works in the literature, e.g. Refs.~\cite{scully2003,rezek2006,rossnagel2014,zhang2014,Bissbort2016minimalistic,Brandner2015}.

Our engine model is sketched in Fig.~\ref{fig:engine}. In its initial configuration, the harmonic working mode is in contact with a hot reservoir, which causes it to thermalize to the average excitation number $\bar{n}_{H}$ at a rate $\kappa$. Radiation pressure then pushes on the piston, which exerts torque on the attached rotor and causes it to spin clockwise. Once the piston passes its bottom turning point, the radiation pressure starts to push against the spin, which would drive the system into pendular motion. To prevent this, the working mode is now brought into contact with a cold reservoir that decreases the average excitation number to $\bar{n}_C$ in order to suppress radiation pressure until the upper turning point is reached and the thermal contact switches from the cold to the hot bath again. This modulated, angle-dependent thermal coupling between the working mode and the two reservoirs is what keeps the wheel spinning clockwise, gaining momentum with every round-trip. Note that this modulation does not conflict with our aim of building an autonomous engine. Indeed, our engine is analogous to a car engine where a crankshaft spins due to contact with an oscillating piston, and a synchronized camshaft controls the opening of the inlet and exhaust valves. In our heat engine, the rotor plays the role of the crankshaft while the modulation of the coupling to the bath plays the role of the camshaft.

\emph{Model.--}Classically, the rotor degree of freedom is described by an angle variable $\varphi \in [0,2\pi)$ and an angular momentum component $L_z$ perpendicular to the plane of rotation. However, in the quantum version of the rotor, the bound spectrum of the angle operator $\hat{\varphi}$ implies a discrete spectrum of $\hat{L}_z$, and Hermitianity must be enforced by imposing periodic boundary conditions~\cite{carruthers1968phase}, i.e. we must work with strictly periodic functions of the angle. In particular, we use the commutation relation $[e^{i\hat{\varphi}},\hat{L}_z]=-\hbar e^{i\hat{\varphi}}$.

The Hamiltonian of the engine in a frame rotating at the mode frequency is given by
\begin{equation}\label{eq:HS}
	\hat{H}_S=\frac{\hat{L}_z^2}{2I}+\hbar g\,\hat{a}^{\dagger}\hat{a}\cos(\hat{\varphi}) ,
\end{equation}
where $I$ is the moment of inertia of the rotor, $g$ is the opto-mechanical coupling strength and $\hat{a}$ is the annihilation operator of the mode. Importantly, the mode frequency does not enter the description as the radiation pressure term is only proportional to the mode occupation $\hat{a}^{\dagger}\hat{a}$. Hence, we are free to match the mode frequency to the requirements of weak linear bath coupling and of the preferred physical implementation.

The Hamiltonian $\hat{H}_S$ may for instance describe an optical Fabry-Perot cavity where one of the mirrors is rigidly attached to the rotor and allowed to move along the $x$ direction~\cite{aspelmeyer2014}. The radiation pressure acting on the mirror will then always try to push the rotor away from the cavity, in proportion to the number of photons in the cavity. Although in this work we are concerned with a theoretical study of the engine, a direct opto-mechanical realization of the engine model can be envisaged given the recent experimental advances in the rotational control of nanorods trapped in a cavity field \cite{kuhn2016full,hoang2016torsional,kuhn2015cavity}. Other physical realizations are also conceivable, where the working mode is not restricted to an optical field mode and where the angular variable might be associated to a phase degree of freedom instead of the mechanical gear depicted in Fig.~\ref{fig:engine}. 

We can now describe the interaction of the mode with its environment, consisting of the hot (H) and cold (C) baths. Specifically, we describe each bath as an ensemble of harmonic oscillators
\begin{equation} \label{eq:HB}
	\hat{H}_T=\int_{-\infty}^{\infty}\dd\omega\,\hbar \omega \hat{b}_T^\dagger(\omega)\hat{b}_T(\omega) ,
\end{equation}
with correlation functions
\begin{eqnarray}\label{eq:noiseCorrel}
	\mean{\hat{b}_T^\dagger(\omega)\hat{b}_T(\omega')}&=&\bar{n}_T\,\delta(\omega-\omega') ,\\
	\mean{\hat{b}_T(\omega)\hat{b}_T^\dagger(\omega')}&=&(\bar{n}_T+1)\,\delta(\omega-\omega')\nonumber ,
\end{eqnarray}
where $T=H,C$ and $\bar{n}_T$ is the associated thermal occupation at the mode frequency. In the Schr\"{o}dinger frame, the interaction between the mode and the two baths is then described by
\begin{equation}\label{eq:HBS}
	\hat{H}_{B-S}=i\hbar\sum_{T=H,C}\int_{-\infty}^{\infty}\dd\omega\,\gamma f_T(\hat{\varphi})\left[\hat{b}_T^\dagger(\omega)\hat{a}-\hat{a}^\dagger\hat{b}_T(\omega) \right] ,
\end{equation}
where we neglect the variation of the coupling constant $\gamma$ and of the thermal occupation $\bar{n}_T$ around the mode frequency. At this point, the only non-standard feature of our model for the baths is the modulation of the coupling via the functions $f_T(\hat{\varphi})$. This synchronicity is what breaks the symmetry, allowing the engine to be pushed harder than it is slowed down. Naturally, this internal clock cannot be used to construct a perpetual machine of the second kind, namely a heat engine that would run on a single heat bath at a fixed temperature. Indeed, the presence of the cold bath is of utmost importance in order to extract heat from the mode and hence lower the radiation pressure. We note that the crucial role of this internal clock for building autonomous machines has also been pointed out recently in the context of solar cells~\cite{Alicki2017}.

When engineering the modulating function $f_H(\varphi)$, it is sufficient to ensure that the mode is coupled strongly with the hot bath in the interval $0<\varphi<\pi$ but only weakly coupled in the interval $\pi<\varphi<2\pi$, and vice versa for $f_C(\varphi)$. For simplicity, we consider the following modulating functions
\begin{equation}\label{eq:modFunc}
	f_H(\hat{\varphi})=\frac{1+\sin(\hat{\varphi})}{2} ,\quad  f_C(\hat{\varphi})=\frac{1-\sin(\hat{\varphi})}{2}.
\end{equation}
This specific choice is motivated by the requirement of working with periodic functions of the angle in the quantum regime. Nevertheless, the results obtained will not be qualitatively affected by a different choice of functions, as long as they alternate with negligible overlap as described above.
A somewhat closer resemblance to the engine of a car can be achieved by working with coupling functions of narrow support, say, on small windows around the angles  $\varphi=0$ (H) and $\pi$ (C). Efficient operation would then require sufficiently fast thermalization within the respective time windows.

To conclude the presentation of our model, we emphasize that while the choice of modulating the coupling rate via the rotor's angular position is genuinely novel, the Langevin equations and the master equation can still be obtained following the original derivations presented in~\cite{gardiner1985}. This will allow us to trace out the bath degrees of freedom and describe their influence in terms of an effective thermalization rate $\kappa =2 \pi \gamma^2$~\footnote{Note that \cite{gardiner1985} uses the opposite convention for $\kappa$ and $\gamma$}. For this treatment to be valid, the latter is assumed to be small compared to the (freely adjustable) mode frequency and the spectral variation in the bath coupling.

\section{Classical Regime}
\emph{Nonlinear stochastic dynamics.--}We start by studying the dynamics of the rotor heat engine in the classical regime. To this end, we consider the classical limit of the quantum Langevin equations for the rotor coordinates and the complex mode amplitude~\cite{gardiner1985},
\begin{eqnarray}\label{eq:eomCla}
	&\!\!\!\!\dd \varphi &= L_z/I\ \dd t,\\
	&\!\!\!\!\dd L_z& = \hbar g |a|^2 \sin(\varphi)\, \dd t - \!\! \sum_{T=H,C} \!\! 2\hbar \sqrt{\kappa \bar{n}_T} f_T' (\varphi)\, {\rm Im} \left( a^{*} \dd w_T \right) , \nonumber\\
	&\!\!\!\!\dd a &= -\left[i g \cos(\varphi)+\kappa (\varphi)/2\right] a\, \dd t - \!\! \sum_{T=H,C} \!\! \sqrt{\kappa \bar{n}_T} f_T(\varphi)\, \dd w_T .\nonumber
\end{eqnarray}
The second and third line are stochastic differential equations in It\^o form~\cite{Gardiner2004}. For a fixed angle $\varphi$, the third line describes the thermalization of the mode with the two baths. Here $w_H$ and $w_C$ are complex Wiener processes, \emph{i.e.} continuous stochastic processes with independent time increments $\dd w_T$ that take complex values following a normal distribution with $|\dd w_T|^2=\dd t$. They correspond to the noise incoming from the baths, with
\begin{equation}
	\kappa(\varphi)=\kappa\left[f_H^2(\varphi)+f_C^2(\varphi)\right]
\end{equation}
the overall decay rate of the mode intensity. For our choice of functions, we have $\kappa/2\leq\kappa(\varphi)\leq\kappa$, such that the mode is always in contact with a thermal bath for any position of the rotor. Note that the noise input also affects the angular momentum of the rotor directly. In fact, the stochastic term in the second line of \eqref{eq:eomCla} can be understood as the classical counterpart of the quantum measurement backaction due to the angle-dependent coupling to the baths. We omit it in the following classical assessment of the engine performance, but we will see later that it accounts for a part of the additional noise in the quantum version of the engine.

In describing the dynamics of the engine, we are interested in the evolution of statistical quantities like the averages and variances of the random variables described in the equations of motion \eqref{eq:eomCla}. However, solving them is not a straightforward matter, especially given the non-linear form of the radiation pressure term driving the angular momentum $L_z$. In order to tackle this problem, we proceed in three steps: (i) reduce the two Wiener processes to a single one (ii) derive the equation of motion for the mode intensity, from which we can simulate efficiently the dynamics (iii) perform an adiabatic elimination of the mode, which will allow us to obtain compact analytical results.

As a first step, we thus make use of the fact that the sum of two independent Wiener processes can be expressed as a single effective Wiener process $w_{\text{eff}}$, namely
\begin{equation} \label{eq:noiseCombine}
	\sum_{T=H,C}\sqrt{\kappa \bar{n}_T} f_T (\varphi)\, \dd w_T=\sqrt{\kappa(\varphi) \bar{n}(\varphi)}\, \dd w_{\text{eff}} .
\end{equation}
In the classical regime, the two baths can thus be reduced to a single bath with an effective thermal occupation modulated by the rotor's position
\begin{equation}\label{eq:nbar}
	\bar{n}(\varphi)=\frac{f_H^2(\varphi)\bar{n}_H+f_C^2(\varphi)\bar{n}_C}{f_H^2(\varphi)+f_C^2(\varphi)} .
\end{equation}
As we shall see later, this intuitive simplification does not generalize to the quantum regime where coherence between different angles may occur.

\begin{figure*}
\begin{center}
\includegraphics[scale=1]{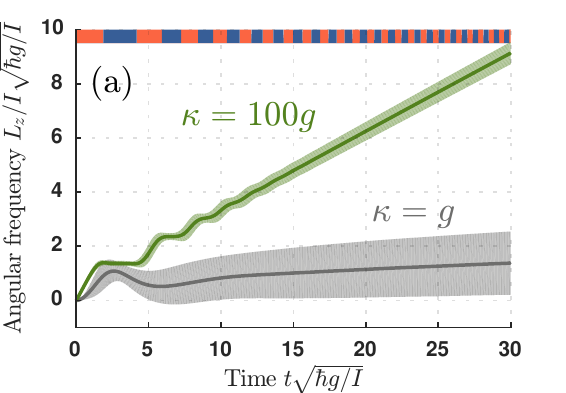}\quad
\includegraphics[scale=1]{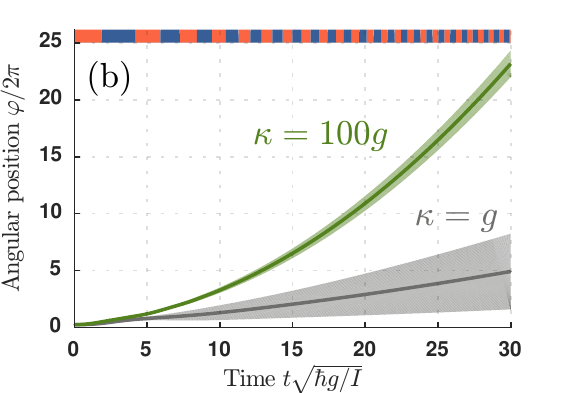}\quad
\includegraphics[scale=1.09]{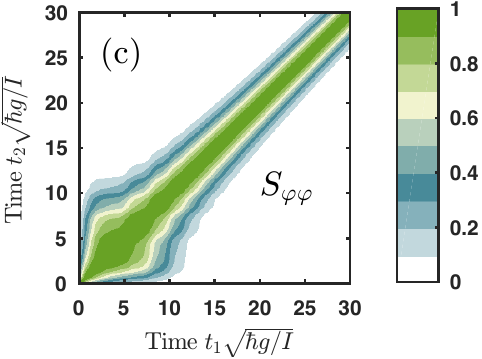}
\end{center}
\caption{Classical simulation of the engine dynamics. Panels (a) and (b) depict, respectively, the angular momentum and the unwrapped angle of the rotor for two exemplary cases of fast ($\kappa=100g$) and slow ($\kappa=g$) thermalization. The lines represent the mean values for a numerical sample over $10^5$ trajectories, the shaded areas cover two standard deviations. Red and blue areas on the top indicate the sign of $\sin \mean{\varphi(t)}$ for $\kappa=100g$, \emph{i.e.}~where the working mode is mostly in contact with the hot and the cold bath. Panel (c) shows the angular two-time correlation function $S_{\varphi\varphi} (t_1,t_2)$ for $\kappa=100g$. $S_{\varphi\varphi} (t_1,t_2)=1$ implies that the angular position at time $t_2$ can be predicted from its value at time $t_1$ (and vice versa) while $S_{\varphi\varphi} (t_1,t_2)=0$ corresponds to uncorrelated angular random variables. All simulations start with the rotor at rest ($L_z(0)=0$,~$\varphi(0)=\pi/2$,~$I=\hbar/g$), and reservoirs at $(\bar{n}_H,\bar{n}_C)=(1,0)$. }\label{fig:classSim}
\end{figure*}

Given that the phase of the mode does not impact the dynamics of the rotor heat engine, the model can be reduced further by solely considering the mode intensity $|a|^2$, whose It\^o stochastic differential equation can be derived from the Fokker-Planck equation~\cite{Risken1996,Gardiner2004},
\begin{eqnarray}\label{eq:ItoInt}
	\dd \varphi &=& (L_z/I) \dd t, \quad \dd L_z = \hbar g |a|^2 \sin(\varphi)\, \dd t \\
	\dd |a|^2 &=& -\kappa(\varphi) \left[|a|^2-\bar{n}(\varphi)\right] \dd t + \sqrt{2 |a|^2\kappa(\varphi)\bar{n}(\varphi)}\,  \dd W \nonumber ,
\end{eqnarray}
where the characteristic frequency of the engine motion is $\sqrt{\hbar g/I}$. Here $W$ is a single real-valued Wiener process ($\dd W^2 = \dd t$), the only remaining source of randomness that enters the classical backaction-free model of the engine. Note that the It\^o calculus used here implies that the dynamical variables $|a|^2(t)$, $L_z(t)$ and $\varphi(t)$ are non-anticipating functions of the noise~\cite{Gardiner2004}, \emph{i.e.} they are independent of the behaviour of the Wiener process $W$ in the future of $t$.

\emph{Numerical simulation.--}Typical results from a numerical Monte Carlo integration of the classical engine model (\ref{eq:ItoInt}) are shown in Fig.~\ref{fig:classSim} for exemplary cases of fast and slow thermalization (see \cite{SupMat} for animated trajectories). Starting from the rotor at rest ($L_z(0)=0$,~$\varphi(0)=\pi/2$), in the fast case the evolution of the average angular momentum $\mean{L_z}$ [green line in panel (a)] shows that the engine is accelerating clockwise. Moreover, the noise (shaded area) is relatively small and does not impact significantly the performance of the engine, which is in contrast to the case of slow thermalization (grey). Looking at the angle variable in panel (b) yields the same conclusion, with an additional subtlety. Indeed, while it is true that the one-dimensional unbounded coordinate $\varphi$ grows with negligible relative noise, the actual angle coordinate of the rotor is defined up to a multiple of $2\pi$. As a consequence, as soon as the standard deviation $\Delta\varphi$ is of order $\pi$, the distribution of angles will essentially appear flat. Physically speaking, this means that one will not be able to infer the exact angle of the rotor from a known value that lies several cycles in the past. This can be seen in Fig.~\ref{fig:classSim}(c) where we plot the two-time correlation function of the periodic angle variable~\cite{fisher1983}
\begin{equation}
	S_{\varphi\varphi}(t_1,t_2)=\frac{R\big[\varphi(t_1)-\varphi(t_2)\big]-R\big[\varphi(t_1)+\varphi(t_2)\big]}{\sqrt{\Big(1-R\big[2\varphi(t_1)\big]\Big)\Big(1-R\big[2\varphi(t_2)\big]\Big)}}
\end{equation}
where $R[\phi]=\mean{\cos\phi}^2+\mean{\sin\phi}^2$. The width of $S_{\varphi\varphi}$ with respect to $|t_1-t_2|$ determines the time window of phase stable motion, before the angular position is completely diffused. However, the fact that the phase stability of the motion is bounded by thermal fluctuations only poses a practical limitation if they are able to stop or reverse the average spinning direction. In other words, the relevant conditions for a steady operation of the engine are that the relative spread $\Delta L_z/\mean{L_z}$ in the angular momentum must be small and that the two-time correlations $S_{\varphi\varphi}$ extend over more than one cycle of rotation.

\emph{Adiabatic elimination.--}The equations for the classical description are exact so far. In order to characterize analytically the engine's operation, we now adiabatically eliminate the mode variable $|a|^2 (t)$. Specifically, we assume that thermalization occurs on a much shorter timescale than the motion of the rotor such that the mode intensity will assume its mean value \eqref{eq:nbar} almost instantaneously for each angle $\varphi(t)$. To be explicit, let us separate the small noise deviations from the mean in the mode intensity, $|a|^2(t) = \bar{n}[\varphi (t)]+\varepsilon_a (t)$, and insert it into the equation of motion \eqref{eq:ItoInt}. We obtain $\dd \varepsilon_a = -\kappa(\varphi) \varepsilon_a \dd t + \sqrt{2\kappa(\varphi)\bar{n}(\varphi)(\bar{n}(\varphi)+\varepsilon_a)}\dd W$, a random variable that contains information of the rotor trajectory integrated over a time scale of $1/\kappa$. At low rotation speed  $L_z/I \ll \kappa$, we can neglect this short-time memory effect and assume that any function of the angle $\varphi(t)$ is non-anticipating for $\varepsilon_a(t)$.

This approximation, which we use to derive the analytical results below, corresponds to the desired regime of operation for the engine (if an external load were attached to the piston). Indeed, it is precisely when the mode is given sufficient time to thermalize with the baths that heat can be extracted to create the required bias in radiation pressure. Optimal performance is thus reached in the limit $\kappa \to \infty$, whereas $\kappa$-dependent corrections are expected to appear for $\kappa \sim \sqrt{\hbar g/I}$ or $\kappa \sim \mean{L_z}/I$.

As the rotor accelerates, it will eventually enter this latter regime where the approximation breaks down and the angular frequency saturates. This situation would be analogous to a car engine that could not follow the opening and closing of its valves given the fast rotation of the camshaft. However, a car engine would rarely operate in such a regime as the targeted velocities under load are kept well below the intrinsic thermalization rates. The saturation can already be seen in Fig. \ref{fig:classSim} (a) for the case $\kappa=g$. Additionally, other effects would start to play a role at high speeds, such as friction of the rotor which is not included in the present model since it is difficult to include rigorously, although it could be done for example following the results of Ref. \cite{stickler2016spatio}.

\begin{figure}
\begin{center}
\includegraphics[scale=1]{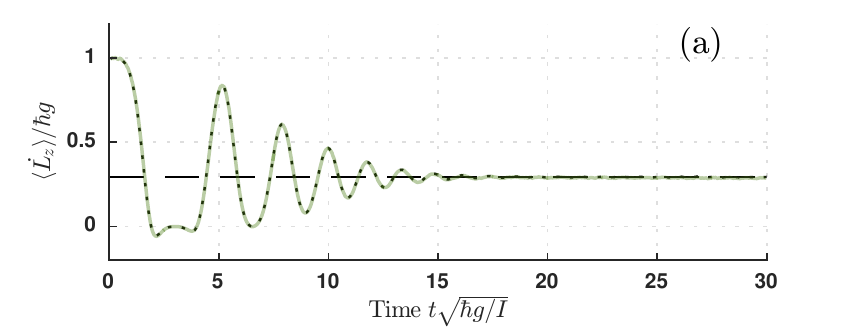}\\
\includegraphics[scale=1]{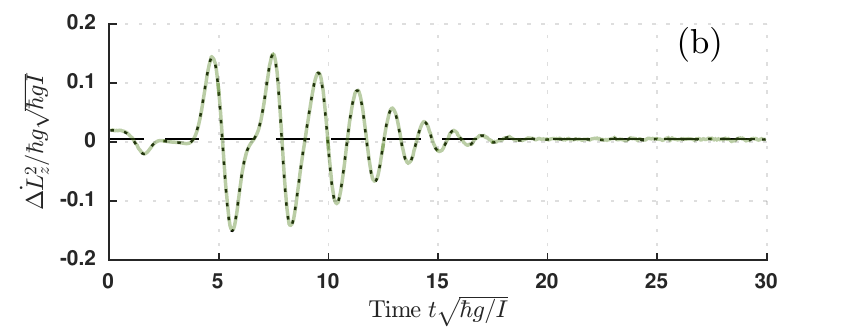}
\end{center}
\caption{Rate of increase of (a) the average angular momentum $\mean{L_z}$ and (b) its variance $\Delta L_z^2$ for the parameters of Fig.\,\ref{fig:classSim}. The green solid line corresponds to the direct computation of the derivative from the simulated dynamics. The dotted and dashed line are obtained from the analytical rates given in Eq.\,\eqref{eq:ratesCla}, respectively before and after taking the limit of free rotation. As expected, the latter description is only valid once the engine has started and the gain in angular momentum within each cycle is sufficiently small. }\label{fig:classRates}
\end{figure}

At this point, we have all the necessary information to derive the rates at which the average angular momentum $\mean{L_z}$ and its variance $\Delta L_z^2$ increase as a function of time. They read
\begin{eqnarray}\label{eq:ratesCla}
	\dot{\mean{L_z}}&=&\hbar g\, \mean{\sin(\varphi)\bar{n}(\varphi)}\,\overset{\text{F.R.}}{\to}\, \hbar g(1-\frac{1}{\sqrt{2}})(\bar{n}_H-\bar{n}_C) ,\\
	\dot{\Delta L_z^2}&=&2\hbar g\, \mean{\sin(\varphi)\bar{n}(\varphi)\Big(L_z-\mean{L_z}+\hbar g\frac{\sin(\varphi)\bar{n}(\varphi)}{\kappa(\varphi)}\Big)}\nonumber\\
	&\!\!\!\!\!\!\!\!\!\overset{\text{F.R.}}{\to}&\! \frac{\hbar^2 g^2}{\kappa}\left[(1-\frac{1}{\sqrt{2}})(\bar{n}_H+\bar{n}_C)^2+\frac{3}{8\sqrt{2}}(\bar{n}_H-\bar{n}_C)^2\right] ,\nonumber
\end{eqnarray}
where $\overset{\text{F.R.}}{\to}$ stands for the limit when the gain in angular momentum per cycle is small enough so that the quantities can be averaged over one round-trip of free rotation, $\mean{g(\varphi)}\overset{\text{F.R.}}{\to}1/{2\pi}\int_0^{2\pi}\dd \varphi \,g(\varphi)$. This yields compact analytical expressions in spite of the nonlinear dynamics (see Fig.\,\ref{fig:classRates} for a comparison with the exact dynamics).

Following our intended goal, the average angular momentum $\mean{L_z}$ is driven in proportion to the difference in thermal occupation of the baths, which in turn drives the average angle coordinate $\mean{\varphi}$. Inevitably, heat also enters the system in the form of noise, limiting the phase stability and accumulating uncertainty in the angle as a function of time. Classically, this happens at finite temperatures even in the absence of driving, $\bar{n}_H-\bar{n}_C = 0$. Indeed, the angular momentum variance $\Delta L_z^2$ acquires contributions from the difference as well as the sum in thermal occupation of the baths. However, it grows linearly in time as does $\mean{L_z}$, which implies that the relative noise $\Delta L_z/\mean{L_z}$ decreases over time. In other words, the rotor behaves as a clock with a steadily improving signal-to-noise ratio.

\begin{figure}
\begin{center}
\includegraphics[width=\columnwidth]{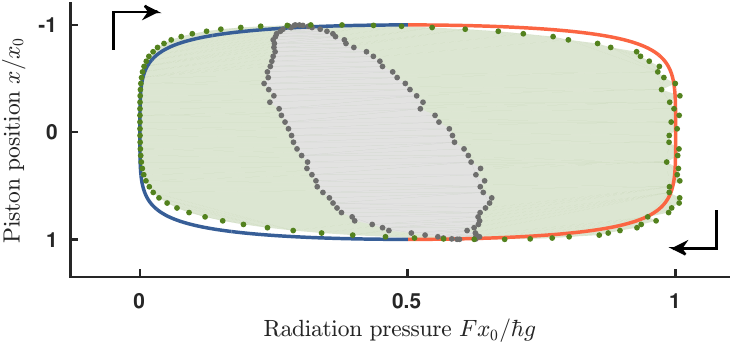}
\end{center}
\caption{Engine cycle in a pressure-volume diagram, where the mode volume grows linearly with the piston position $x$. The solid line represents the ideal cycle running clockwise and producing $\mathcal{W}_{\rm cyc}$, where the cavity has time to thermalize $|a|^2 \to \bar{n}(\varphi)$. The color indicates the sign of $\sin \phi$, i.e.~the dominant bath coupling. The markers show the mean radiation pressure in $100$ bins of $\varphi \mod 2\pi$, sampled from $10^6$ trajectories evolved to the time $gt=30$,  with the parameters of the fast (outer cycle) and slow (inner cycle) cases chosen as in Fig.~\ref{fig:classSim}.}\label{fig:workCycle}
\end{figure}

\emph{Work output.--} By looking at the dynamics, we have shown that the classical version of the rotor heat engine is achieving its goal of extracting useful directional motion from the heat transfer between the baths. How well it performs its task can be further quantified in terms of energy flows. In each engine cycle, the working mode is extracting a positive net amount of useful rotational energy by pushing via the radiation pressure force onto the piston. The latter plays the role of a flywheel \cite{Levy2016} that stores this energy, e.g.~for later extraction by external loads. The corresponding mechanical work generated along the cycle is given by
\begin{equation}\label{eq:deltaW}
 \delta \mathcal{W} = F\dd x = \hbar g |a|^2 \sin (\varphi) \dd \varphi.
\end{equation}
where $x = -x_0 \cos(\varphi)$ is the vertical piston position which determines the volume of the working mode. The pressure is given by the force that pushes downwards, i.e.~in the positive $x$-direction. It is proportional to the mode intensity~\cite{law1995,aspelmeyer2014}, $F = \hbar g |a|^2/x_0$, as follows from the optomechanical potential in \eqref{eq:HS}. In the ideal case of fast thermalisation where $|a|^2 \to \bar{n}(\varphi)$, we can integrate over one cycle to obtain the upper bound $\mathcal{W}_{\rm cyc} = \hbar g\pi(2-\sqrt{2})(\bar{n}_H - \bar{n}_C)$ on the mean work output per cycle. Figure~\ref{fig:workCycle} illustrates the engine cycle in a pressure-volume diagram, comparing the ideal cycle (solid line, clockwise rotation) to a snapshot of $10^6$ simulated trajectories based on the same parameters as in Fig.~\ref{fig:classSim}. The area enclosed by the data points yields the average work output, upper bounded by the ideal case (solid line). The data for $\kappa=100g$ (green outer cycle) follows the ideal curve closely and outputs almost maximum work (98\%), whereas the data for $\kappa=g$ (grey inner cycle) performs significantly worse (27\%) since the working mode already lacks time to thermalize with each bath. This confirms our previous observation in Fig.~\ref{fig:classSim}(a) that the mechanical output of the engine deteriorates in the saturated regime of fast rotation.

A direct consequence of the autonomous engine design is that the cycle duration is fluctuating and decreasing over time, as opposed to externally driven Otto cycles where the clock reference is provided by the machine operator~\cite{zhang2014,Campo2016}. It will prove expedient to measure the engine's work performance in terms of the mean output power
\begin{eqnarray}\label{eq:workPower}
 \mathcal{P}_{\mathcal{W}} =\mean{\frac{\delta \mathcal{W}}{\dd t}} &=& \frac{\hbar g}{I}\mean{|a|^2 \sin (\varphi) L_z }\\
 &\approx& \frac{\hbar g}{I}\mean{\bar{n}(\varphi) \sin (\varphi) L_z } \,\overset{\text{F.R.}}{\to}\, \frac{\mathcal{W}_{\rm cyc}}{2\pi}\frac{\mean{L_z}}{I}, \nonumber
\end{eqnarray}
i.e.~the rate at which work is performed on the piston \cite{Deffner2013}. It is a function of the angular frequency in the same way the horsepower of a car engine is a function of the number of revolutions per minute (RPM). In the limit of free rotation and fast thermalization, the mean output power is simply given by the work per cycle divided by the average period of rotation.

Notice that the output power gives the average rate at which the kinetic energy $L_z^2/2I$ of the rotor increases due to radiation pressure, as described by the backaction-free dynamics \eqref{eq:ItoInt}. This means that a finite output power does not guarantee that the engine is spinning in a fixed direction, a prerequisite for the generated motion to be useful. Using \eqref{eq:workPower} as a figure of merit for the engine performance is thus only meaningful in combination with $\mean{L_z}^2 \gtrsim \mean{L_z^2}$.

\emph{Heat input.--} Having identified the work output $\delta \mathcal{W}$ in Eq.\eqref{eq:deltaW} as the change of the working mode energy caused by the moving piston, we can identify heat as the energy change resulting from a change in mode occupation $|a|^2$. In accordance with the first law of thermodynamics, the energy change $\dd E=\dd (\hbar \omega(\varphi) |a|^2)$ due to heat transfer is thus
\begin{equation}
	\delta \mathcal{Q} = \dd E +\delta \mathcal{W} = \hbar\omega(\varphi) \,\dd |a|^2 ,
\end{equation}
where $\omega (\varphi) = \omega_0 + g \cos(\varphi)$ is the mode's resonance frequency modulated as a function of the piston position. Note that we consider the standard regime of cavity optomechanics~\cite{law1995,aspelmeyer2014} where this modulation of the mode's boundary condition, which gives rise to the optomechanical potential in \eqref{eq:HS}, is accounted for to first order in $\Delta\omega=2g \ll \omega_0$.

To compute the efficiency of the engine, we must distinguish between the heat input $\delta \mathcal{Q}_{H}$ from the hot reservoir and the output into the cold. The change in the mode occupation $\dd |a|^2 $ corresponding to its thermalization with the hot bath is given by the respective dissipator plus noise,
\begin{equation}\label{eq:dHn}
	\dd_H |a|^2=-\kappa f^2_H(\varphi)\left[|a|^2-\bar{n}_H\right]\,\dd t\ (+\mathrm{\ noise})  ,
\end{equation}
where the noise term will average out in the following. In an idealized scenario of clearly separated work and heat strokes~\cite{zhang2014,rossnagel2014}, where the mode is allowed to thermalize with each reservoir separately and is entirely isolated from them when performing work, the mean input of the hot bath would add up to $\hbar\omega_H(\bar{n}_H - \bar{n}_C)$ per cycle, corresponding to the hot thermalization stroke at a frequency $\omega_H$~\cite{zhang2014}. In contrast, our choice of overlapping coupling functions $f_C(\varphi)$ and $f_H(\varphi)$ leads to a greater average heat consumption since the simultaneous interaction with both reservoirs implies a balance of heat flows, where the hot bath input \eqref{eq:dHn} must counteract losses into the cold bath in order to maintain the average occupation $\bar{n}(\varphi)$, similar to the Stirling engine of Ref.~\cite{rossnagel2016single}. Hence the mode never reaches the highest mean value $\bar{n}_H$ except at $\varphi=\pi/2$. We quantify the average heat consumption as a function of time in terms of the input power
\begin{eqnarray} \label{eq:heatPower}
	\mathcal{P}_{H} =\mean{\frac{\delta \mathcal{Q}_{H}}{\dd t}} &=& \hbar\kappa \mean{\omega(\varphi) f^2_H(\varphi)(\bar{n}_H-|a|^2 )}\\
	&\approx& \hbar\kappa \mean{\omega(\varphi) f^2_H(\varphi)(\bar{n}_H-\bar{n}(\varphi))} \nonumber \\
	&& \,\overset{\text{F.R.}}{\to}\, \hbar\omega_0\kappa\frac{\sqrt{2}-5/4}{4}(\bar{n}_H-\bar{n}_C) .\nonumber
\end{eqnarray}
Here the second line is a good approximation in the regime of fast thermalization, contrary to a scenario of separated strokes of hot and cold thermalization.

\begin{figure}
\begin{center}
\includegraphics[width=\columnwidth]{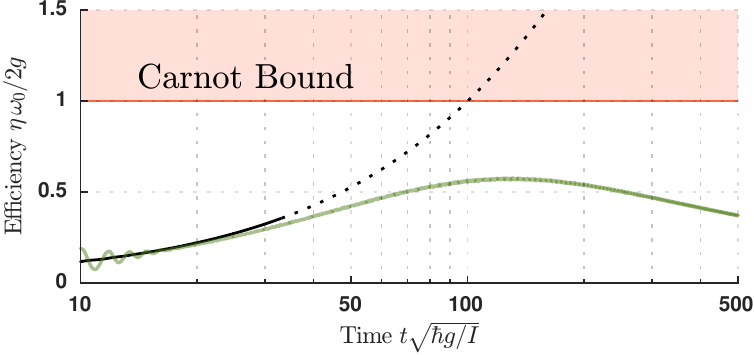}
\end{center}
\caption{Efficiency of the engine in units of $2g/\omega_0$ for the parameters of Fig.\,\ref{fig:classSim}. The green solid line corresponds to the direct computation of the efficiency from the simulated dynamics, using the expressions \eqref{eq:workPower} and \eqref{eq:heatPower} before the adiabatic elimination and the free-rotation limit. The black line is obtained from the analytical result given in Eq.\,\eqref{eq:eff} and is plotted in the regime of operation (solid) and in the saturated regime $\mean{L_z}/I\kappa \geq 10^{-1}$ (dotted). At short times, the efficiency oscillates as the free-rotation limit is not valid, while in the long-time limit the efficiency deviates from \eqref{eq:eff} due to the mode not thermalizing fast enough.}\label{fig:effC}
\end{figure}

\emph{Efficiency.--} The thermal efficiency of the engine in the non-stationary case without external load is
\begin{equation}\label{eq:eff}
	\eta = \frac{\mathcal{P}_{\mathcal{W}}}{\mathcal{P}_{H}}\,\overset{\text{F.R.}}{\to}\,\frac{2g}{\omega_0}\frac{\mean{L_z}}{I\kappa}\frac{2-\sqrt{2}}{\sqrt{2}-5/4} .
\end{equation}
Our autonomous design implies that the efficiency is a function of the parameter $\mean{L_z}/I\kappa$. In particular, the faster the engine spins, the better its efficiency, as shown in Fig.\,\ref{fig:effC}. This behavior is valid up to the regime where $\mean{L_z}/I\sim\kappa$, in which case the adiabatic elimination does not hold anymore, as discussed previously. In fact, this dependence is a feature shared with actual car engines, for which the efficiency typically follows a bell-shaped curve as a function of the RPM \footnote{This dependency of the engine efficiency with RPM has led the automotive industry to come up with innovative solutions such as autonomously varying the opening of the valves as a function of the RPM, with technologies known as Variable Valve Timing and Lift Electronic Control (VTEC) or Variable Valve Timing with intelligence (VVTi).}. In our case, higher efficiencies could in principle be obtained by adapting the profiles $f_H(\varphi)$, $f_C(\varphi)$ and/or the thermalization rate $\kappa$ as a function of the angular frequency $L_z/I$.

The efficiency \eqref{eq:eff} is inherently small, being proportional to both $2g/\omega_0\ll 1$ and $\mean{L_z}/I\kappa\ll 1$. However, given the absence of explicit temperatures, a natural question that arises is whether our engine is bounded by the Carnot efficiency $\eta_\mathrm{Carnot}=1-T_C/T_H$ with the temperatures $T_H > T_C$ associated to the two baths. To answer this question, we first note that given the modulation of the mode frequency $\omega(\varphi)$, the corresponding thermal occupation 
\begin{equation}\label{eq:therm_occupation}
\bar{n}_T(\omega)=(e^{\hbar \omega/k_B T}-1)^{-1}
\end{equation}
is also slightly fluctuating along the cycle, which we have neglected in our model. In order for the engine to produce work, the lowest possible hot occupation number $\bar{n}_H(\omega_0+g)$ must be greater than the highest possible cold occupation $\bar{n}_C(\omega_0-g)$ number. Inserting this into Eq. \eqref{eq:therm_occupation} leads to the condition 
\begin{equation}
\frac{\omega_0-g}{T_C} > \frac{\omega_0+g}{T_H}
\end{equation}
which in turn implies
\begin{equation}
\eta_\mathrm{Carnot} > 1-\frac{\omega_0-g}{\omega_0+g}=\frac{2 g}{\omega_0} +\mathcal{O}\left[ \left(\frac{g}{\omega_0}\right)^2 \right].
\end{equation}
As shown in Fig.\,\ref{fig:effC}, the efficiency of the engine remains below this bound in both the regime of operation $\mean{L_z}/I\kappa\ll 1$, described by Eq.~\eqref{eq:eff}, as well as in the saturated regime where the efficiency starts to drop as expected.

\section{Quantum Regime} \label{sec:quantum}
\emph{Master equation.--}The master equation governing the dynamics of the quantum heat engine is given by~\cite{gardiner1985}
\begin{eqnarray}\label{ME}
\dot{\hat{\rho}}&=&-\frac{i}{\hbar}\left[\hat{H}_S,\hat{\rho}\right]\\
&+&\sum_{T=H,C}\kappa(\bar{n}_T+1)D\left[f_T(\hat{\varphi})\hat{a}\right]\hat{\rho}+\kappa\bar{n}_T D\left[f_T(\hat{\varphi})\hat{a}^{\dagger}\right]\hat{\rho} ,\nonumber
\end{eqnarray}
where $D[\hat{\mathcal{O}}]\hat{\rho}=\hat{\mathcal{O}}\hat{\rho} \hat{\mathcal{O}}^\dg-\frac{1}{2}\left\{\hat{\mathcal{O}}^\dg\hat{\mathcal{O}},\hat{\rho}\right\}$ is the Lindblad superoperator~\cite{lindblad1976}. 

As stated before, the mode frequency does not enter the description as the radiation pressure term is only proportional to the mode occupation. This means that we can match the frequency to the weak-coupling condition required for the validity of the master equation. In Appendix~\ref{app:ME}, we derive this master equation using a weak-coupling approach, and we check that the Lindblad terms are consistent with a thermodynamic interpretation of the engine dynamics in terms of entropy flows \cite{Levy2014,trushechkin,stockburger}.

In contrast to the classical regime, the dissipative coupling of the mode to the reservoirs cannot be reduced to an effective single-bath term. This is a manifestation of quantum coherence in the angle coordinate $\varphi$, since off-diagonal matrix elements $\bra{n,\varphi}\hat{\rho}\ket{n,\varphi'}$ are influenced by both reservoirs simultaneously. In fact, the Lindblad dissipators will lead to decoherence in the angle coordinate, as the exchange of photons with the reservoirs reveals information about the piston position, i.e.~constitutes a coarse angle measurement by the environment.

\begin{figure*}
\begin{center}
\includegraphics[scale=1]{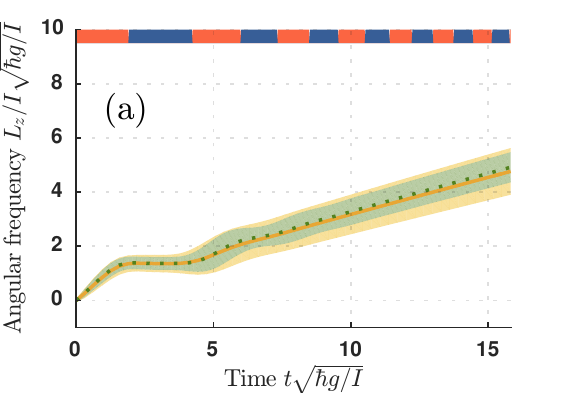}\quad
\includegraphics[scale=1]{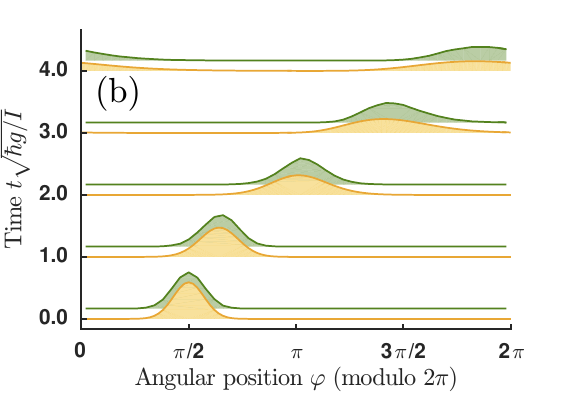}\quad
\includegraphics[scale=1]{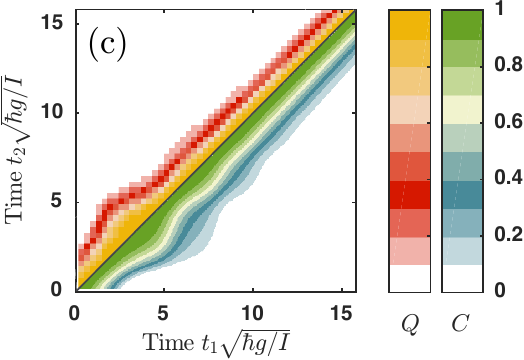}
\\
\includegraphics[scale=1]{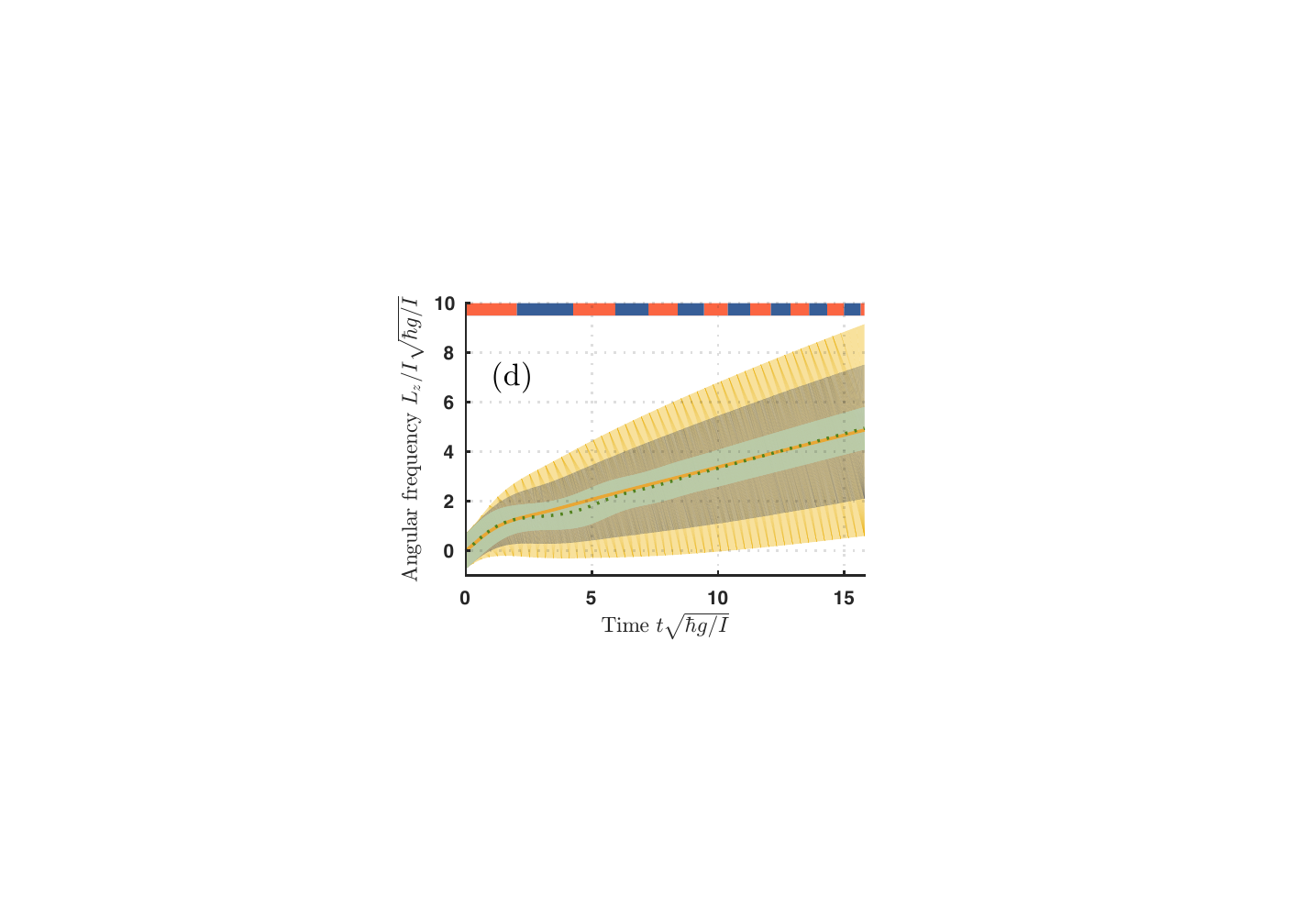}\quad
\includegraphics[scale=1]{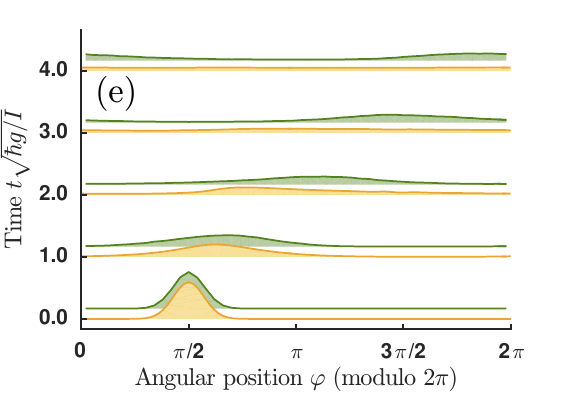}\quad
\includegraphics[scale=1]{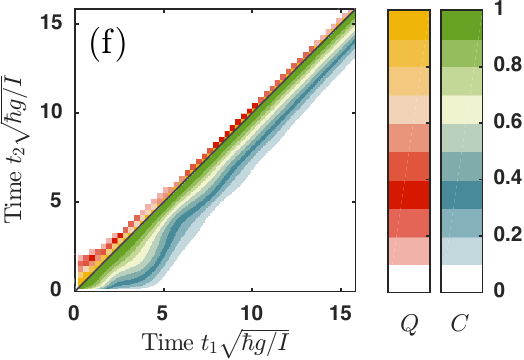}
\end{center}
\caption{Simulation of the engine dynamics for $I g / \hbar = 100 k$ (first row) and $I g /\hbar = k$ (second row). All the results obtained from the quantum model (in yellow) are compared to the classical backaction-free case (in green). Panels (a) and (d) depict the angular momentum while (b) and (c) show the evolution of the angular distribution. Panels (c) and (f) show the angular two-time correlation function $S_{\varphi\varphi} (t_1,t_2)$, which we symmetrize in the quantum regime to obtain a real quantity (note that the classical version is always real and symmetric). The grey area in (d) shows the uncertainty predicted by a classical model including backaction noise. The engine parameters are ($k=10$,~$\mu=\pi/2$,~$\kappa=10^{3/2}\sqrt{\hbar g/I}$).}\label{fig:QSim}
\end{figure*}

\emph{Initialization.--}For the rotor to start spinning in the right direction, we initialize its angular position in the region $\varphi\in[0,\pi]$ where the mode is predominantly coupled to the hot bath. This implies that any convex combination of energy eigenstates -- such as a thermal distribution -- should be avoided, since in this case the rotor is completely delocalized. Alternatively, one could start with an initial displacement in momentum. However, this would require an external energy source, just like the battery of a starting car engine, which we avoid in our autonomous model.

Here we select an initial pure state given by the periodic von Mises wavefunction
\begin{equation}\label{vonMises}
\psi_i(\varphi)=\frac{e^{k\cos(\varphi-\mu)}}{\sqrt{2\pi I_0(2k)}},
\end{equation}
where $I_0(2k)$ is a modified Bessel function. With this choice, the angular position of the rotor is localized around a mean value $\mu$ with a spread determined by the parameter $k$. For a sufficiently localized angle, i.e. for $k\gg 1$, the angle distribution is approximately Gaussian, with a standard deviation $1/\sqrt{2k} \ll \pi/2$. 
The corresponding angular momentum distribution covers a spectrum of order $\sqrt{k}$ quanta around its zero average. No further initialization is required, and any net rotation of the rotor will come entirely from the engine dynamics.

Ideally, the rotor will evolve in such a way that its average angular momentum $\mean{\hat{L}_z}$ increases with time, while the relative spread $\Delta L_z/\mean{\hat{L}_z}$ remains small. In contrast to the classical model, however, quantum mechanics will impose additional constraints for the performance of the engine. In particular, for the rotor to start spinning, the initial acceleration of the engine must exceed the free dispersion of the rotor. This is roughly the case if the initial spread in kinetic energy, $\Delta E(0) \approx \hbar^2 k/2I$, is small compared to the gain $\hbar g \bar{n}_H$ in potential energy during the first half-cycle, $I g \bar{n}_H/\hbar \gg k$. The larger $I g \bar{n}_H/\hbar$ is, the more the quantum rotor approaches classical behaviour and the less it is prone to free dispersion.

\emph{Numerical simulations.--}
We employ two numerical methods to simulate the dynamics of the quantum engine. One is direct integration using the QuTip package in Python over a truncated Hilbert space \cite{johansson2013qutip}. Specifically, we restrict the rotor Hilbert space to angular momentum quantum numbers $m_{\min} \leq m \leq m_{\max}$, with suitably chosen bounds to cover the occupied spectrum at all times. The thermal mode is allowed to have at most 8 excitations, which is sufficient given that we operate with $(\bar{n}_H,\bar{n}_C) = (1,0)$ throughout. Higher reservoir temperatures are computationally expensive but do not provide further insight into the engine's operation.  
We choose $\bar{n}_C=0$ for simplicity, but of course this should be interpreted as a negligibly small mean photon number at appropriate mode frequencies and temperatures, $\hbar \omega_0 \gg k_B T_C$. We consider $0 < \bar{n}_C \ll 1$ in Appendix~\ref{app:ME} for an assessment of the system entropy, to avoid divergences that would occur at exactly zero temperature. 

The other method is a stochastic sampling of the master equation \eqref{ME} in terms of piecewise deterministic jump trajectories \cite{Breuer2002}. We use direct integration to solve for the dynamics of the engine, and stochastic sampling to explore values of the engine parameters that lead to a good performance of the engine.

\emph{Quantum vs. Classical.--} The results of the numerical simulations for both the classical backaction-free model and the quantum model are summarized in Fig. \ref{fig:QSim} for two different choices of parameters. We have selected values of the moment of inertia $I$ and the coupling strength $g$ such that $I g/\hbar \gg k$ in one case, while $I g/\hbar = k$ in the other case. Additionally, these values are set such that the classical dynamics are essentially unchanged in both cases, allowing us to showcase how properties that are irrelevant classically become meaningful in a quantum setting.

As discussed previously in reference to Eq. \eqref{vonMises}, due to the uncertainty principle, it is impossible to perfectly localize both the angular position and angular momentum of a quantum rotor. In order to make a fair comparison and mimic this in the classical case, we initialize the rotor's angular position and angular momentum in a Gaussian probability distribution, with mean $\pi/2$ and standard deviation $1/\sqrt{2k}$ for the angular position, and mean $0$ and standard deviation $\sqrt{k/2}$ for the angular momentum. This corresponds to the Gaussian approximation of the von Mises distribution~\eqref{vonMises}, which holds for $k\gg 1$. This initialization of the classical rotor results in a free dispersion of the angle coordinate over time similar to the quantum case, and so it allows us to distinguish quantum effects that arise due to initialization from effects that originate from the rotor's interaction with the working mode.

As seen in Fig. \ref{fig:QSim}(a), the quantum engine shows almost identical behaviour for the angular frequency as in the classical case when $I g/\hbar \gg k$, as expected. However, for $I g/\hbar = k$ in panel (d), the quantum model yields a much larger variance around the mean value, even though the latter still increases steadily in time. We can also examine the distribution of the angular position of the rotor as a function of time. Here we see that there is significantly less broadening when $I g/\hbar \gg k$ in panel (b), whereas the angular distribution is almost flat before it completes one revolution for $I g/\hbar =k$ in panel (e). Classically, this is explained by different spreads in angular frequency for a given spread in angular momentum. In the quantum case, however, the additional noise contributions not only broaden the angle distribution further, but also impact the phase stability of the rotor engine. This is shown in panels (c) and (f), where the symmetric two-time correlation function $S_{\varphi\varphi}(t_1,t_2)$ is plotted for the quantum (upper triangle) and classical (lower triangle) cases. While the phase stability of the classical engine survives in the regime $I g/\hbar = k$, correlations drop almost instantaneously in the quantum case.

\emph{Classical backaction.--} In fact, in the two regimes we have explored, the amount of noise in the quantum case is strictly larger than in the classical one. This additional uncertainty arises due to the combined effect of measurement backaction noise and of vacuum fluctuations contributing to the noise input of the hot and the cold bath. While the latter is a genuine quantum feature, backaction noise can be accounted for in a classical engine model. Specifically, the difference to the backaction-free model \eqref{eq:ItoInt} is an additional noise term in the equation for the angular momentum variable,
\begin{eqnarray}\label{eq:dLz_classBA}
 \dd L_z &=& \hbar g |a|^2 \sin(\varphi)\, \dd t  \\
 &&- \hbar \sqrt{2\kappa|a|^2 \left\{ \bar{n}_C \left[f'_C (\varphi) \right]^2 + \bar{n}_H \left[f'_H (\varphi) \right]^2 \right\}} \dd U. \nonumber 
\end{eqnarray}
Here, $\dd U$ stands for the increment of a second, independent Wiener process, $\mean{\dd U \dd W}=0$ (See Appendix~\ref{app:BA} for a derivation based on the classical Langevin equations \eqref{eq:eomCla}). For an exemplary comparison, the uncertainty predicted by this model is evaluated and depicted as the grey-shaded area in Fig.\,\ref{fig:QSim}(d). It accounts for about half of the excess noise of the quantum model for the specific low-temperature parameters considered here. At high temperatures, $\bar{n}_C, \bar{n}_H \gg 1$, we expect a good match with the quantum prediction, even in the regime of low inertia 
where the additional uncertainty would mainly come from the backaction noise.

\begin{figure}
\begin{center}
\includegraphics[scale=1]{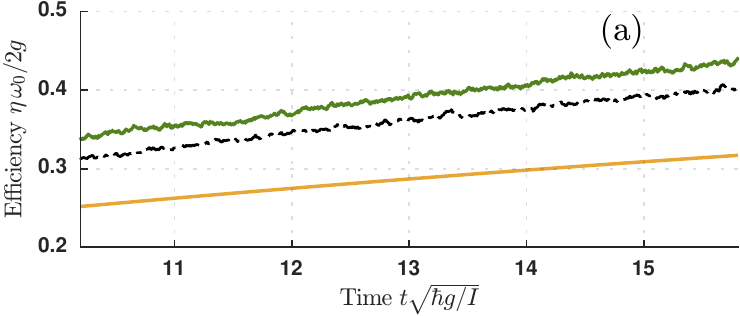}\\
\includegraphics[scale=1]{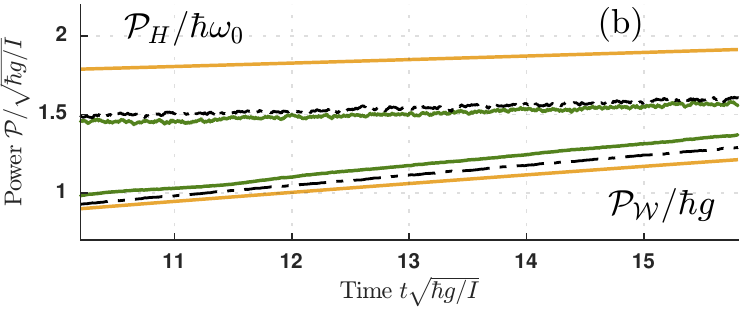}
\end{center}
\caption{Performance of the engine in the regime $I g /\hbar = k$ of Fig.\,\ref{fig:QSim} where quantum effects play a role. The quantum model (in yellow) is compared to the classical results averaged over $5\times 10^6$ trajectories with (black dashed) and without backaction noise (green). (a) Efficiency in units of $2g/\omega_0$. (b) Heat input power (top three lines) and work output power (bottom three lines).}\label{fig:effQ}
\end{figure}

\emph{Quantum efficiency.--} In Figure \ref{fig:effQ}, we compare the quantum and the classical version of the engine in terms of the thermal efficiency \eqref{eq:eff}. It is defined in terms of the mean work output power \eqref{eq:workPower} and heat input power \eqref{eq:heatPower}, which carry over to the quantum model if we replace the rotor and mode coordinates by the respective operators and symmetrize the product of $\hat{L}_z$ and $\hat{\varphi}$-dependent terms for hermiticity. 
The input power follows from the master equation \eqref{ME} if we consider only the dissipator to the hot bath for the time derivative of the cavity energy. 
We define the power $\mathcal{P}_{C}$ of the heat flow dumped into the cold bath in the same way,
\begin{eqnarray} \label{eq:Pin_Q}
 \mathcal{P}_{H,C} &=& \hbar\kappa \mean{ \omega(\hat{\varphi}) f_{H,C}^2 (\hat{\varphi}) (\bar{n}_{H,C} - \hat{a}^\dagger\hat{a}) }. 
\end{eqnarray}
The mechanical output power reads as
\begin{equation}
  \mathcal{P}_{\mathcal{W}} = \frac{\hbar g}{2I} \mean{ \hat{a}^\dagger\hat{a} \left\{ \sin(\hat{\varphi}),\hat{L}_z \right\} },
\end{equation}
which also corresponds to the change of the rotor's kinetic energy due to radiation pressure (excluding the contribution of backaction noise).
In total, a dynamical version of the first law \cite{Alicki1979} can be formulated for the time derivative of the work mode energy in the quantum case,
\begin{equation}
  \partial_t \mean{ \hbar \omega(\hat{\varphi}) \hat{a}^\dagger\hat{a} } = \mathcal{P}_{H} + \mathcal{P}_{C} - \mathcal{P}_{\mathcal{W}}.
\end{equation}
Note that, in order to make sense of the work output stored in the rotational motion, we define the energy balance with respect to the work mode subsystem excluding the rotor degree of freedom. Hence we do not account for the increase of its kinetic energy due to backaction, as given by the power
\begin{eqnarray} \label{eq:PBA}
  \mathcal{P}_{\mathcal{B}} &=& \frac{\rm d}{{\rm d} t} \mean{\frac{\hat{L}_z^2}{2I}} -\mathcal{P}_{\mathcal{W}} \\
  &=& \frac{\hbar^2 \kappa}{4I} \mean{\left[ \frac{\bar{n}_H+\bar{n}_C}{2} (2 \hat{a}^\dg \hat{a} + 1) + \hat{a}^\dg \hat{a} \right] \cos^2 (\hat{\varphi}) } \geq 0. \nonumber
\end{eqnarray}
While we have seen that it can be responsible for a significant noise contribution to the motion of low-inertia rotors, $\mathcal{P}_{\mathcal{B}}$ will be much smaller than the direct heat flows \eqref{eq:Pin_Q} in and out of the work mode for $\omega_0 \gg \hbar/I$. Hence we can safely operate with $\mathcal{P}_{H}$ as the input power to assess the quantum efficiency.

We also remark that the intrinsic definition of $\mathcal{P}_{\mathcal{W}}$, which reflects the growth in kinetic energy of the rotor, will in general overestimate the amount of work that could be extracted by an external load in the low-inertia quantum regime. Indeed, the large uncertainty around the average angular momentum in Fig.~\ref{fig:QSim}(d) indicates a significant difference between the rotor's average kinetic energy $\mean{\hat{L}_z^2}/2I$ and the energy $\mean{\hat{L}_z}^2/2I$ associated to its average directional motion. Formally, the latter could be extracted by applying a unitary momentum displacement operator, leaving behind passive energy in the form of momentum noise \cite{Allahverdyan2004,Niedenzu2017}. The distinction turns irrelevant as the momentum signal-to-noise improves in the high-inertia regime.

We use the low-inertia parameter regime of before ($Ig/\hbar = k$) for a pronounced difference between the quantum and the classical models. The yellow line in Fig.~\ref{fig:effQ}(a) represents the quantum result, which is systematically below the classical results with (black dashed) and without (green) backaction. In fact, as shown in Fig.~\ref{fig:effQ}(b), the classical version of the engine is consistently extracting more work from less input heat. The reason lies in the additional vacuum fluctuations acting on the working mode and deteriorating correlations between the number operator $\hat{n}$ and the rotor observable $\sin(\hat{\varphi})$. The sign of the latter encodes whether the rotor is predominantly coupled to the hot or the cold reservoir, and it thus gets correlated with $\hat{n}$ as the engine operates. These correlations enter the mean input \eqref{eq:heatPower} and output \eqref{eq:workPower} power and are affected by both the backaction noise and the vacuum fluctuations, as is reflected by the lower efficiencies in Fig.~\ref{fig:effQ} compared to the classical backaction-free result. The classical and quantum efficiencies converge in the regime of high inertia ($Ig/\hbar \gg k$) and high temperature ($\bar{n}_{C} \gg 1$) where neither backaction nor vacuum noise play a role.

Overall, the quantum heat engine operates best when it approaches the classical limit of a large moment of inertia together with a large coupling strength. Free dispersion, vacuum fluctuations, and backaction noise do not directly affect the average mechanical output, but they can ruin the phase stability of the rotor and are thus problematic for the proper functioning of the engine. Our results showcase the importance of studying the actual dynamics of heat engines and of addressing countermeasures to quantum sources of noise.

\section{Conclusion}
Inspired by actual piston engines, we have proposed an autonomous rotor heat engine described by standard Hamiltonians. By solving the underlying equations of motion in the transient regime where the rotor accelerates from rest, we have shown analytically that the engine functions as desired in the classical regime, while identifying the best regimes of operation. We have also explored the role of quantum effects, and our results show that, in the case of our engine, they mainly give rise to additional noise in the motion of the rotor. It is a relevant question whether other quantum effects such as entanglement and coherence can lead to a better performance compared to a fully classical version of the engine. To this end, our rotor heat engine provides a suitable testbed for the various notions and concepts that have been put forward in the context of quantum thermodynamics. For example, the study of Otto-cycle-type of modulating functions, as well as quantum-engineered initial states of the rotor, could open new room for improvement of the engine.

With our results, our aim is to generate new theoretical insights into how thermal energy can be converted autonomously into useful mechanical motion, and to understand the role that quantum mechanics plays in thermal machines. With respect to an implementation of the engine, the technological challenge is to simultaneously build a system where the thermalization rate is fast compared to the angular frequency of the rotor, where the dissipation of the rotor is negligible on the timescale of operation of the engine, and where the frequency of the mode is significantly larger than the coupling to the environment. Typical high-Q cavities are well suited to implement the working mode~\cite{aspelmeyer2014}. As for the rotor degree of freedom, it is possible to consider non-mechanical systems where a phase variable plays the role of the angular position, such as the flux in electric circuits based on Josephson junctions.

Finally, we note that an analogy can be drawn between our engine and the one-dimensional motion of a particle in a periodic potential generated by, say, a standing-wave cavity field \cite{ritsch2013cold}. In this context, the acceleration of the rotor can also be understood as the reverse of a Sisyphus cooling scheme, where the rotor runs effectively more down- than uphill on the potential energy curve \cite{dalibard1985dressed,dalibard1989laser}.

\begin{acknowledgments}
We thank Paul Skrzypczyk, Dario Poletti, Colin Teo, and Marc-Antoine Lemonde for fruitful discussions. This research is supported by the Singapore Ministry of Education through the Academic Research Fund Tier 3 (Grant No. MOE2012-T3-1-009); by the National Research Foundation, Prime Minister’s Office, Singapore, through the Competitive Research Programme (Award No. NRF-CRP12-2013-03); and by both above-mentioned source, under the Research Centres of Excellence programme.
\end{acknowledgments}

\appendix

\section{Weak-coupling derivation of the master equation} \label{app:ME}

Here we rederive the master equation \eqref{ME}, which was stated following the Gardiner-Collett derivation \cite{gardiner1985} applied to each bath. Following the standard Born-Markov secular approach \cite{CarmichaelStatMeth1,Breuer2002}, we switch to the interaction picture with respect to the free system and bath Hamiltonians \eqref{eq:HS} and \eqref{eq:HB},  and we rewrite the interaction Hamiltonian \eqref{eq:HBS} as $\hat{H}^{I} (t) =i\hbar \sum_{T=H,C}\left[ \hat{A}_T(t) \hat{B}^\dagger_T (t) - \mathrm{H.c.} \right]$ with
\begin{eqnarray}\label{eq:AB_Int}
	\hat{A}_T (t) &=& e^{i\hat{H}_S t/\hbar} \hat{a} f_T (\hat{\varphi}) e^{-i\hat{H}_S t/\hbar}, \\
	\hat{B}_T (t) &=& \int_{0}^{\infty} \dd \omega \, \Gamma(\omega) \hat{b}_T (\omega) e^{-i(\omega-\omega_0)t}. \nonumber
\end{eqnarray}
Contrary to the simplified Gardiner-Collett treatment, we still keep the restriction to positive physical frequencies and account for the frequency dependence of the coupling rate density $\Gamma(\omega)$ at this point.
Notice that the secular approximation is already implied as only resonant interaction terms are considered; for this we must assume that the modulation of the bare mode frequency by radiation pressure is negligible, $g \ll \omega_0$. 

In the Born approximation, we describe the two independent baths by stationary Gibbs states $\hat{\sigma}_T$ in the formal solution for the time evolution of the reduced system state $\hat{\rho}^{I}(t)$, 
\begin{eqnarray}
 \dot{\hat{\rho}}^{I} (t) &=& -\int_0^t \frac{\dd t'}{\hbar^2} \tr_{H,C} \left\{ \left[ \hat{H}^{I} (t), \left[ \hat{H}^{I} (t'), \hat{\rho}^{I}(t') \hat{\sigma}_H \hat{\sigma}_C \right]\right] \right\}. \nonumber \\ \label{eq:ME_raw}
\end{eqnarray}
Using the fact that the bath modes at different frequencies are uncorrelated, i.e.~the noise correlation functions \eqref{eq:noiseCorrel} with a frequency-dependent occupation $\bar{n}_T(\omega)$, the non-vanishing bath correlation functions are given by
\begin{eqnarray}
\mean{\hat{B}_T^\dagger (t)\hat{B}_T (t') } &=& \int_0^\infty \dd \omega\, \Gamma^2(\omega) \bar{n}_T (\omega) e^{i(\omega-\omega_0)(t-t')}\nonumber \\
&\approx& \frac{\kappa}{2} \bar{n}_T (\omega_0) \delta (t-t'), \\
\mean{\hat{B}_T (t)\hat{B}^\dagger_T (t') } &\approx& \frac{\kappa}{2} [\bar{n}_T (\omega_0) +1] \delta (t-t'). \nonumber
\end{eqnarray}
In the Markov approximation step, we omit the Lamb shift, and we restrict to the coarse-grained time scales of the relevant system dynamics. We are allowed to do so if the latter is much slower than the inverse correlation times $1/t_{H,C}$ of the two reservoirs, which in turn must be much smaller than $\omega_0$. 
Specifically, we consider the engine to operate in a regime where the effective thermalization rate $\kappa$ marks the fastest time scale in the evolution of the system state. Hence the Markov approximation requires that $\kappa \ll 1/t_{H,C} \ll \omega_0$. For free-space radiation, the typical correlation time depends on the temperature, $t_{H,C} \sim \hbar/k_B T_{H,C}$, which is typically much shorter than achievable decay times of cavity modes at realistic temperatures $T_{H,C}$. 

The Markov assumption amounts to using the noise correlation functions \eqref{eq:noiseCorrel} with a constant $\gamma$ and $\bar{n}_T = \bar{n}_T (\omega_0)$ in the Gardiner-Collett approach, and it allows us to approximate \eqref{eq:ME_raw} by setting the upper integral bound to infinity and replacing $\hat{\rho}^{I}(t') \to  \hat{\rho}^{I}(t)$ in the integrand. The resulting time-local master equation reduces to
\begin{eqnarray}
 \dot{\hat{\rho}}^{I} (t) &\approx& \kappa \sum_{T=H,C} \left\{ (\bar{n}_T + 1) D\left[ \hat{A}_T (t) \right] \hat{\rho}^{I} (t) \right. \nonumber \\ 
 && \left. + \bar{n}_T D\left[ \hat{A}^\dagger_T (t) \right] \hat{\rho}^{I} (t) \right\}, \label{eq:ME_raw2}
\end{eqnarray}
which is equivalent to \eqref{ME} in the Schr\"{o}dinger frame (rotating at the bare mode frequency $\omega_0$). 

In general, a heuristic description of system-bath couplings to different temperatures by simply adding the associated Lindblad dissipators to the master equation may lead to a violation of the second law of thermodynamics \cite{Levy2014}. This does not happen here, as we have checked by computing the von Neumann entropy $\mathcal{S}(t) = - \tr (\rho \ln \rho)$ of the engine over time. The intrinsic entropy production rate is obtained by taking the time derivative and subtracting the contributions associated to the heat flows in and out of the two reservoirs, 
\begin{equation} \label{eq:dSint}
 \dot{\mathcal{S}}_{\rm int} = \frac{{\rm d} \mathcal{S}}{{\rm d} t} - \frac{\mathcal{P}_{H}}{k_B T_H} - \frac{\mathcal{P}_{C}}{k_B T_C}.
\end{equation}
Assuming that the bare mode frequency $\omega_0$ exceeds the rotation frequency quantum $\hbar/I$ by several orders of magnitude, we omit the additional backaction heating term \eqref{eq:PBA} that would act directly on the rotor.
Negativity of the entropy production rate would indicate a violation of the second law \cite{Alicki1979}. Notice however that, in the absence of external loads or rotor dissipation, the combined system of work mode and rotor will not evolve towards a stationary (Gibbs) state. 
In fact, we rather observe that the entropy of the engine grows steadily over time as the engine spins up and more and more energy is stored in the rotor. 

\begin{figure}
\begin{center}
\includegraphics[scale=1]{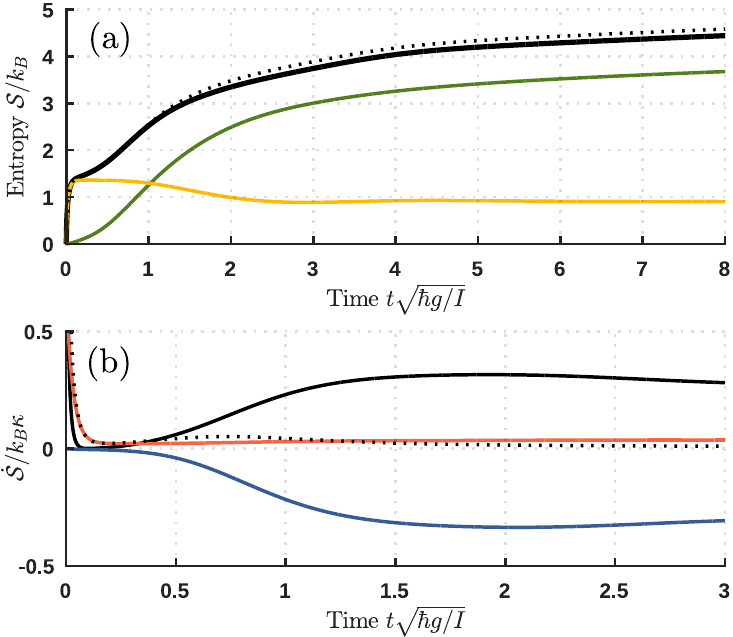}
\end{center}
\caption{(a) Entropies of the quantum engine in the regime $I g /\hbar = k$ of Fig.\,\ref{fig:QSim}, with $\bar{n}_C = 10^{-3}$. The von Neumann entropy of the engine state in units of $k_B$ (black) is compared to the entropies of the reduced work mode and rotor states (yellow and green), and to the sum of both (black dotted). (b) Rate of entropy change (black dotted) versus entropy flows in and out of the engine (red and blue) in units of $k_B \kappa$ at short times. The black solid line shows the intrinsic entropy production rate \eqref{eq:dSint}. It is always positive, with a minimum value of $7.7 \times 10^{-4} k_B \kappa$ at $t = 0.1 \sqrt{I/\hbar g}$. }\label{fig:Entropy}
\end{figure}

To evaluate \eqref{eq:dSint}, finite temperatures are now explicitly required, in particular $\bar{n}_C > 0$. We started from the parameter regime explored in Sec.~\ref{sec:quantum} and varied $\bar{n}_C=10^{-6}, 10^{-3}, 10^{-1}$. In all cases, the intrinsic entropy production rate was consistently positive. Fig.~\ref{fig:Entropy} illustrates the case of $ \bar{n}_C = 10^{-3}$, whose results for the relevant engine observables discussed in the main text are practically indistinguishable from the zero-temperature idealization. Panel (a) depicts the von Neumann entropy $\mathcal{S}(t)$ of the engine state (black) and of the reduced states of work mode and rotor (yellow and green). The sum of the latter (dotted) evolves slightly above former, indicating some correlation between rotor and work mode. In (b) we plot the intrinsic entropy production rate (black solid) and compare it to the individual terms in \eqref{eq:dSint}, viz.~the overall change ${\rm d}\mathcal{S}/{\rm d}t$ (dotted), and the entropy flows $\mathcal{P}_{H}/k_B T_H$ and $\mathcal{P}_{C}/k_B T_C$ from the two baths (red and blue).

\section{Classical engine model with backaction} \label{app:BA}

We can simplify the classical Langevin equations \eqref{eq:eomCla} by expressing the working mode in terms of its action-angle variables (number and phase), i.e.~$a = a_r + ia_i =  \sqrt{n}\exp (i\theta)$ with $n=|a|^2$ and $\theta = \arctan (a_i/a_r)$. Noting that each complex Wiener process is a linear combination of two independent real-valued Wiener processes, $\dd w_T = (\dd X_T + i\dd Y_T )/\sqrt{2}$ with $\dd X_T^2 = \dd Y_T^2 = \dd t$, we obtain
\begin{eqnarray} \label{eq:eomCla-1}
 \dd \varphi &=& (L_z/I) \dd t \\
 \dd L_z &=& \hbar g n \sin (\varphi) \dd t \nonumber \\
 && - \hbar \sum_{T} \sqrt{2 \kappa n \bar{n}_T} f_T' (\varphi) \left[ \cos(\theta) \dd Y_T - \sin(\theta) \dd X_T \right], \nonumber \\
 \dd n &=& -\kappa(\varphi) \left[ n - \bar{n} (\varphi) \right] \dd t \nonumber \\
 && - \sum_{T} \sqrt{2 \kappa n \bar{n}_T} f_T (\varphi) \left[ \cos(\theta) \dd X_T + \sin(\theta) \dd Y_T \right]. \nonumber
\end{eqnarray}
The last equation is obtained using the It\^o rule, $\dd n = a \dd a^{*} + a^{*} \dd a + \dd a \dd a^{*}$. We omit the equation for the phase $\theta$ as it turns out to be irrelevant. In fact we notice that the square-bracketed terms in \eqref{eq:eomCla-1} define a rotation of $\dd X_T$ and $\dd Y_T$ to two new mutually independent Wiener processes given the non-anticipating phase angle $\theta$, 
\begin{equation}
 \left( \begin{array}{c} \dd W_T \\ \dd V_T \end{array} \right) = \left( \begin{array}{cc} \cos(\theta) & \sin(\theta) \\ -\sin(\theta) & \cos(\theta) \end{array} \right) \left( \begin{array}{c} \dd X_T \\ \dd Y_T \end{array} \right),
\end{equation}
where $\mean{\dd W_T \dd V_T} = 0$ and $\dd W_T^2 = \dd V_T^2 = \dd t$. Hence the rotor dynamics does not depend on $\theta$. 

Next we simplify further by combining the noise inputs of the hot and the cold bath in the same way as in  \eqref{eq:noiseCombine}. This leaves us with only two independent Wiener processes $U, W$, and
\begin{eqnarray} \label{eq:eomCla-2}
 \dd L_z &=& \hbar g n \sin (\varphi) \dd t \\
 && - \hbar \sqrt{2 \kappa n \left\{ \bar{n}_H \left[ f_H' (\varphi) \right]^2 + \bar{n}_C \left[ f_C' (\varphi) \right]^2\right\}}  \dd U , \nonumber \\
 \dd n &=& -\kappa(\varphi) \left[ n - \bar{n} (\varphi) \right] \dd t \nonumber \\
 && - \sqrt{2 \kappa n \left[ \bar{n}_H f_H^2 (\varphi) + \bar{n}_C f_C^2 (\varphi) \right]} \dd W. \nonumber
\end{eqnarray}
The second equation for the mode intensity corresponds to the one in \eqref{eq:ItoInt}. The angular momentum equation, which contains the backaction noise, equals \eqref{eq:dLz_classBA}. For our choice of coupling functions, it further reduces to
\begin{equation}
 \dd L_z = \hbar g n \sin (\varphi) \dd t - \hbar \cos (\varphi) \sqrt{\kappa n \frac{\bar{n}_H + \bar{n}_C}{2}}  \dd U ,
\end{equation}
as used in the classical simulation underlying the grey-shaded results of Fig.~\ref{fig:QSim}(d). We used Euler steps for the numerical integration with appropriately small time steps $\Delta t$ and normally distributed random numbers $\Delta U,\Delta W$ of variance $\Delta t$ around zero. Strong convergence of single trajectories can be improved by using Milstein's method \cite{Kloeden1992}, which we employed for Fig.~\ref{fig:workCycle}.

\end{document}